\documentclass[jocse]{jocseart}

\usepackage{booktabs} 

\usepackage{multicol}
\usepackage{multirow}
\usepackage{color,colortbl}
\definecolor{mygray}{gray}{0.9}
\usepackage{url}
\usepackage{subfig}

\setcopyright{jocsecopyright}

\jocseDOI{10.22369/issn.2153-4136/x/x/x }

\begin{document}
\title[Trends in Scientific Computing Training Delivered by a HPC Center]{Trends in Demand, Growth, and Breadth in Scientific Computing Training Delivered by a High-Performance Computing Center}
\author{Ramses van Zon}
\email{rzon@scinet.utoronto.ca}
\author{Marcelo Ponce}
\orcid{0000-0001-5850-7240}
\email{mponce@scinet.utoronto.ca}
\author{Erik Spence}
\email{ejspence@scinet.utoronto.ca}
\author{Daniel Gruner}
\email{dgruner@scinet.utoronto.ca}
\affiliation{%
  \institution{SciNet HPC Consortium, University of Toronto}
  \streetaddress{661 University Ave., Suite 1140}
  \city{Toronto}
  \state{Ontario}
  \postcode{M5G 1M1}
  \country{Canada}}

\renewcommand\shortauthors{Van Zon, R. et al.}

\begin{abstract}

We analyze the changes in the training and educational
efforts of the SciNet HPC Consortium, a Canadian academic High
Performance Computing center,
in the areas of Scientific Computing and High-Performance Computing,
over the last six years.  Initially, SciNet offered isolated
training events on how to use HPC systems and write parallel
code, but the training program now consists of a broad range of
workshops and courses that users can take toward certificates in
scientific computing, data science, or high-performance computing.
Using data on enrollment, attendence, and certificate numbers from
SciNet's education website, used by almost 1800 users so far, we 
extract trends on the growth, demand, and breadth of SciNet's training
program. Among the results are a steady overall growth, a sharp and
steady increase in the demand for data science training, and a wider
participation of 'non-traditional' computing disciplines, which has
motivated an increasingly broad spectrum of training offerings.  Of
interest is also that many of the training initiatives have evolved
into courses that can be taken as part of the graduate curriculum at
the University of Toronto.

\end{abstract}

%
%

%
%


\maketitle

\section{Motivation}

Not that long ago, many larger scientific computations for academic
research were performed on clusters built by researchers using
commodity components strung together by a commodity network,
i.e. Beowulf clusters.  The research groups building these clusters
were also the ones using it.  The knowledge of how to build, program
and use these systems were transmitted from one graduate student to
the next, with perhaps the briefest of instructions posted on a website.  Every system was a bit different, which did not
matter too much, as the knowledge to use the system was already
in-house, and the systems were maintained by a member of the research
group that happened to have affinity with the technology.

This era of self-built clusters was driven by the need for more
computational resources and faster computations than a single
workstation could provide, which is reflected in the name of the
field, High Performance Computing (HPC).  It was made possible by the
availability of relatively cheap commodity hardware and technical
knowledge in the research groups.  This practice of self-built,
maintained, and documented systems often took place in traditionally
highly technical areas such as engineering and the physical
sciences (physics, astronomy, chemistry).

However, the demand for computational resources in academic research
has not stopped increasing, and is not exclusive to the physical
sciences.  Part of this demand is now driven by the increase of data
availability in many disciplines and industries: bioinformatics,
health sciences, social science, digital humanities, commerce,
astronomy, high energy physics, etc.

With the increased demand for scientific computational resources
build-your-own clusters were no longer sufficient.  A select few
researchers in some countries had access to large national systems,
but most researchers did not. To solve this issue, many researchers
started using shared clusters, first within their department, then
their own university or HPC consortia in which several universities
collaborated, and finally nationally and cross-nationally available
shared computing resources
(XSEDE \cite{XSEDE}, Compute Canada \cite{ComputeCanada}, PRACE \cite{PRACE}).
But this brought about a second issue, that the knowledge on how to use these
systems no longer resided within the research groups. This especially
put 'non-traditional' fields at a disadvantage: while they now had
access to great computational resources, they lacked the institutional
knowledge required to use these as effectively as possible and missed
the opportunity to develop this knowledge within their own research
groups.

This computational knowledge gap is the motivation for training 
provided by experts that are situated at the centres that provide the
HPC resources.  Larger computing centres \cite{PSCedu, EPCCedu, BSCedu, website:SciNet-Training}  
have usually engaged in these kinds of training events. But such endeavours
do not automatically tie in with universities' educational systems.
This implies that participants of such training events do not get
formal recognition of the skills and knowledge they learned, unless an
explicit mechanism for this is put in place, such as badges and
certificates.

This paper focuses on SciNet's training efforts. SciNet is the HPC centre at the University of Toronto. It
hosts some of the largest academic supercomputers in Canada. 
Since SciNet's operations started in 2009
\cite{1742-6596-256-1-012026},  courses have been taught on
scientific technical computing, high performance computing, and data
analysis for the Toronto-area research community.

As will be detailed below, what started as small-scale training
sessions on parallel programming has grown into a large, successful
program consisting of seminars, workshops, courses, and summer
schools, delivered by computational science experts. Graduate courses
are given in partnership with other departments, and all events are
part of a program in which participants work towards certificates in
scientific computing, high-performance computing, or data science.  In
this paper, we use the (anonymized) enrollment data of SciNet's
online learning system, which includes attendance records for almost
1800 students over the last seven years, augmented by information on
field of study and gender, to get a picture of the growth in amount
and depth of training in general scientific computing, as well as
trends in training in data science and high performance computing\footnote{The curated and anonymized data can be requested
from the authors for academic research purposes.}.

\section{Training Formats}

The training at SciNet takes place in various different formats.  These
will be briefly describe here, including their intended usage and
their advantages and disadvantages.

Common to all training events is a substantial on-line presence which
supports the learning and administration of the program. SciNet's
training and education site\footnote{\texttt{courses.scinet.utoronto.ca}}
contains lecture videos, slides, links, forums and other electronic
material, freely publicly available and organized by course
\cite{website:SciNet-Edu}.

On the site, users of the SciNet facilities can log in with their
SciNet account, while students that are not users must be
assigned a temporary account. Logging in is not required to access the
content, but is required to enroll in courses, to take tests, to
submit assignments (for the graduate-style courses), and to track
progress towards earning certificates.

All of SciNet's training is free for anyone working in academia.

\subsection{Seminars}

Seminars are short, one-hour sessions. Sometimes they are about a
technical topic, sometimes they are a research presentation.  The
format of seminars may not be ideal for knowledge transfer and
training, but it is a good way learn about something new.
Furthermore, many of these seminars happen at the monthly SciNet
User Group (``SNUG'') meetings, which are an opportunity for 
SciNet users to come together and exchange experiences.

\subsection{Workshops}

Workshops are usually half a day to one day long, and focus on a very
specific topic. Examples of topics are ``Parallel I/O'', ``Relational
Database Basics'', ``Intro to the Linux Shell'', ``Intro to HPC'', and
``Intro to Neural Networks''.  Such workshops are given a few times
throughout the year, and typically have a hands-on component.

The annual summer school (further described below) consists of a
carefully selected collection of these kinds of workshops.

SciNet also provides occasional workshops for other organizations,
such as the Fields Institute, Creative Destruction Lab, and the Chemical BioPhysics
symposium in Toronto.

\subsection{Graduate-style courses}

The graduate-style courses share a common approach which is focussed
on the practical application of presented materials.  These courses
typically have two lectures per week of one hour each. In addition, each
week, students are given a programming assignment, with a due date one
week after, and feedback is given to the students in the following
week. These assignments are designed to help absorb the course
material. The average of the assignments also make up the final grade.
To further support the students' learning, there are office hours,
online forums, and instructor email support.

Initially, these courses ran for four weeks at a time, a
format that coincided with the ``mini'' or ``modular'' courses given
by the Physics Department and the Astrophysics and Astronomy
Department of the University of Toronto.  Topics for these mini-courses included ``Scientific
Software Development and Design'', ``Numerical Tools for Physical
Scientists'', ``High Performance Scientific Computing'',
``Introduction to Programming with Python'', ``Numerical Computing
with Python'', ``Advanced Parallel Scientific Computing'',
``Introduction to Machine Learning'', and ``Introduction to Neural
Networks''.

Some of these graduate-style mini-courses have grown into full-fledged,
term-long graduate courses. The process of creating these recognized
graduate courses is described in more detail in section
\ref{sec:forcredit} below.

\subsection{Summer schools}

The annual one-week long summer school is a flagship training event
for graduate students, undergraduate students, postdocs and
researchers who are engaged in compute intensive research.
SciNet's first summer school was given in 2009 and was called a
``Parallel Scientific Computing'' workshop.  As the program in
table~\ref{table:summerschool2009} shows, it was heavily focussed on
HPC, parallel programming, and applications in astrophysics.

\begin{table}[t]
\begin{tabular}{p{.45\columnwidth}|p{.45\columnwidth}}
\hline
\rowcolor{mygray}\multicolumn{2}{c}{\textit{First day}} \\
\hline
        \multicolumn{2}{c}{Welcome and Introduction to Parallel Scientific Computing}      \\
        \multicolumn{2}{c}{Introduction to OpenMP with brief C tutorial}  \\
\hline
\rowcolor{mygray}\multicolumn{2}{c}{\textit{Second day}}        \\
\hline
        \multicolumn{2}{c}{Introduction to OpenMP, continued}       \\
        \multicolumn{2}{c}{Introduction to MPI}       \\
\hline
\rowcolor{mygray}\multicolumn{2}{c}{\textit{Third day}}        \\
\hline
        \multicolumn{2}{c}{Map Making}       \\
        \multicolumn{2}{c}{Compressible Hydrodynamics}       \\
\hline
\rowcolor{mygray}\multicolumn{2}{c}{\textit{Fourth day}}       \\
\hline
        \multicolumn{2}{c}{OpenMP N-Body}       \\
        \multicolumn{2}{c}{MPI N-Body}       \\
\hline
\rowcolor{mygray}\multicolumn{2}{c}{\textit{Fifth day}} \\
\hline
        \multicolumn{2}{c}{CUDA N-Body}       \\
        \multicolumn{2}{c}{C-Blocks and Erlang}       \\
        \multicolumn{2}{c}{Advanced Topics: Additional Resources,}\\
	\multicolumn{2}{c}{Performance tools and exotic architectures}	\\
\hline\hline
\end{tabular}
\caption{SciNet's first summer school in 2009 focussed on Parallel
  Scientific Computing and placed emphasis on scientific applications
  such as in astrophysics.	
}
\label{table:summerschool2009}
\end{table}

\begin{table*}
\begin{tabular}{p{.275\textwidth}|p{.25\textwidth}|p{.425\textwidth}}
        \hline\hline
        HPC Stream      &       Data Science Stream     &       BioInformatics/Medical Stream   \\
        \hline\hline
        \rowcolor{mygray}\multicolumn{3}{c}{\textit{First day}} \\
        \hline
        \multicolumn{3}{c}{Welcome and Introduction to HPC and SciNet}  \\
        Programming Clusters with MPI  &       Introduction to Linux Shell        &       Python for MRI Analysis \\
        \hline
        \rowcolor{mygray}\multicolumn{3}{c}{\textit{Second day}}  \\
        \hline
        \multirow{2}{.275\textwidth}{Programming Clusters with MPI (cont.)}    &       Introduction to R     &       Image Analysis at Scale \\
		&       Introduction to Python  &       HCP with HPC: Surface Based Neuroimaging Analysis \\
        \hline
        \rowcolor{mygray}\multicolumn{3}{c}{\textit{Third day}}  \\
        \hline
        \multirow{2}{.275\textwidth}{Programming GPUs with CUDA}    &       Parallel Python         &       PLINK   \\
		&       Machine Learning with Python    &               Next Generation Sequencing      \\
        \hline
        \rowcolor{mygray}\multicolumn{3}{c}{\textit{Fourth day}}  \\
        \hline
        \multirow{2}{0.275\textwidth}{Programming GPUs with CUDA (cont.)}  &       Neural Networks with Python     &       RNASeq Analysis   \\
		&       Scientific Visualization Suites   &     R for MRI Analysis      \\
        \hline
        \rowcolor{mygray}\multicolumn{3}{c}{\textit{Fifth day}}  \\
        \hline
        \multirow{2}{.275\textwidth}{Shared Memory Parallel Programming with OpenMP}	&       Debugging, Profiling    &       Public Datasets for Neuroscience        \\
        	&       Bring-Your-Own-Code Lab        &       Biomedical Hacking      \\
        \hline\hline
\end{tabular}
\caption{SciNet's latest (and largest) summer school, held in June
  2018.  This summer school had three parallel streams: the
  traditional High-Performance Computing, one on Data Science and a
  stream on BioInformatics/Medical applications, which was added in
  2017.
	Details of the courses covered in the school can be found in the Summer School website \cite{ss2018}.
	}
\label{table:summerschool2018}
\end{table*}

These days, SciNet's summer school is part of the Compute Ontario
Summer School on Scientific and High Performance Computing. Held
geographically in the west, centre and east of the province of
Ontario, the summer school provides attendees with the opportunity to
learn and share knowledge and experience in high performance and
technical computing on modern HPC platforms.  The central edition is
the continuation of the SciNet summer school.  Not only is the school
organized in a wider context, its program has expanded as well. As
table~\ref{table:summerschool2018} shows, in 2018 there were three
streams in the Toronto edition, and a wide variety of topics, from
shell programming to data science, machine learning and neural networks,
biomedical computing, and, still, parallel programming.

This type of event not only benefits the students and participants of
the summer school, but also enables collaborations between departments
and consortia, as part of the training was delivered in partnership
with colleagues from SHARCNET \cite{sharcnet} and the Centre for Addiction and Mental
Health \cite{CAMH}.

SciNet participates also in the International HPC Summer School
\cite{ihpcss}, sending a few instructors and 10 students to this
competitive one-week program every year.

\subsection{Guest Instructors}

In the capacity of ``guest instructors'', SciNet also delivers a 7-week
module in an undergraduate ``Research Projects Course''
from the Department of Physics at the University of Toronto.
Topics include an introduction to High Performance and Advanced Research Computing,
Data Science, and Scientific Visualization.

SciNet also occasionally provides guest lectures in other courses.

\section{Certificates and Credits}

\subsection{Certificate Programs}

Since December 2012, SciNet has offered recognition to attendees of its
training events in the form of SciNet Certificates
\cite{website:SciNet-Certificates}.  Requirements for these
certificates are based on the number of credit-hours of SciNet courses
a student has successfully completed.  For a short course (typically a
day long or shorter, without homework), a lecture hour counts as one
credit hour; for a long course with homework due between sessions, a
lecture hour counts as 1.5 credit hours.

There are currently three certificate programs, with the following descriptions:

\textbf{Certificate in Scientific Computing:} Scientific computing is
now an integral part of the scientific endeavour. It is an
interdisciplinary field that combines computer science, software
development, physical sciences and numerical mathematics. This
certificate indicates that the holder has successfully completed at
least 36 credit-hours worth of SciNet courses in general
scientific computing topics.

\textbf{Certificate in High Performance Computing:} High Performance
Computing, or supercomputing, is using the largest available computers
to tackle big problems that would otherwise be intractable. Such
computational power is needed in a wide range of fields, from
bioinformatics to astronomy, and big data analytics. Since the largest
available computers have a parallel architecture, using and
programming high performance computing applications requires a
specialized skill set. Those earning this certificate 
have  completed  at least 36 credit-hours of
SciNet courses in high performance computing topics.

\textbf{Certificate in Data Science:} The SciNet Certificate in Data
Science attests that the holder has successfully taken at
least 36 credit-hours of data science-related SciNet courses.  The
latest certificate launched, it is indeed one of the fastest
growing in popularity, clearly displaying the growing interest
in data-science-related computational fields, such as artificial
intelligence and deep learning. 
  
\subsection{For-Credit Graduate Courses}
\label{sec:forcredit}

By partnering with different institutions at the University of
Toronto, many SciNet courses have been consolidated into recognized
graduate courses that students enrolled in Masters and Doctorate
programs can take as part of their graduate curriculum.  So far,
SciNet has started three graduate courses recognized at the University
of Toronto: PHY1610 ``Scientific Computing for Physicists'' in
partnership with the Department of Physics, MSC1090 ``Introduction to
Computational Biostatistics with R" in partnership with the Institute
of Medical Science, and EES1137
``Quantitative Applications for Data Analysis'' in partnership with
the Department of Physical and Environmental Sciences at the University
of Toronto at Scarborough. These courses are in principle open to
students of other departments as well, and indeed attract students
from Physics, Chemistry, Astrophysics, Ecology and Evolutionary
Biology, Engineering, Computer Science, and others.

Some of the shorter graduate-style courses are still taught as well,
and are recognized by a subset of the departments at the university as
``mini'' or ``modular'' courses.

In fact, the physics graduate course started out as a collection of
three such modular courses, on ``Scientific Software Development and
Design'', ``Numerical Tools for Physical Scientists'', and ``High
Performance Scientific Computing'', respectively.  These three modules
were recognized by the Physics, Astrophysics, and Chemistry
Departments, and were subsequently merged into a single, one-term
course with a Physics designation.  This meant the course was now
listed in the graduate curriculum and drew a larger audience. 
Similar tracks were followed to establish the other full-term graduate
courses, but under different partner departments.

To avoid a growing teaching burden, teaching assistant support is
provided by the partner department.  The course instructors are still
SciNet analysts, now hired as sessional lecturers for the purpose of
these courses.

These for-credit courses follow the same format as our other
graduate-style courses, with assignment-based learning and evaluation,
with on-line support in terms of forums, email, and availability of
materials including lecture recordings.  There is no final exam for
these courses, although for some courses a mid-term exam is set.

It should be noted that designing a course in scientific computing for
students in a non-traditional field such as medicine and biology poses
its own unique problems, which are discussed in more detail
elsewhere \cite{CSEgap-inpress}.

\section{Enrollment Data}

The start of the certificate program in 2012 required a more
comprehensive online registration and learning management system than
SciNet's previous Drupal-based site could provide.
The replacement system is based on
ATutor \cite{atutor}, an open-source web-based learning management
system. This system was augmented with a few in-house-developed
modules for event management, certificate programs, and integration
with the LDAP authentication server used for SciNet's computing
resources.  In addition to the LDAP authentication for users with computing
accounts, there are temporary accounts,
which are authenticated separately through a local database.

The site keeps track of all courses in which users are enrolled, and
which of those courses have been completed.  Every course is also
categorized.  For clarity, in this paper, a restricted set of four
categories is used: ``High Performance Computing'', ``Scientific
Computing'', ``Data Science'' and ``Seminar''.

Each course consists of a set of 'events', e.g. lectures, meetings, or
workshops, which in total determine the length of the course and when
it was given.  Attendance of events by people without an account on
the site is allowed, and is tracked by entering the number of
'anonymous attendees' for each event.
The system also records whether users have earned a certificate
in Scientific Computing, High Performance Computing or Data Science,
and the date when they obtained it.

Unfortunately, the system was not setup to gather all the information
needed to, for instance, investigate the distribution of training
demand over different genders and different fields.  To be able to
investigate this, it was necessary to assign genders and fields to
users of the site. For anonymous attendees, this assignment is
impossible, but for the users with accounts on the system, 
these attributes were reconstructed as well as possible given the
information that was in the system.

The gender assignment for accounts whose gender was not known was
performed by checking first names against an online database that
returns the most likely gender (\url{https://genderize.io}).  In the end,
the data set of 1776 users was determined to contain 1047 males and
567 females, leaving 162 unknown.

The assignment of fields of study was done in a variety of ways.  For
users with accounts on computing resources, the research group is
known and for most groups their field of study is known.  
Temporary users were often asked for a description of what they do,
from which the research field could be deduced.  For the graduate
courses, the field of the user's 'host' department was used. Sometimes
the email address of a user revealed his or her field. The assignment of
the field of study was the most laborious part of the data analysis,
only made possible by using a restricted set of categories of fields
of study: 'engineering', 'physics', 'chemistry', 'earth science',
'computer science', 'mathematics', 'medicine', 'biology', 'economics',
'humanities' and 'social science'.  In the end, 1577 of the 1776 user
accounts on the site could be assigned a field of study.

The data covers the period from 2012 to July
2018. Thus, the statistics for 2018 do not constitute a full
year. Furthermore, the 2012 data was imported from the older courses
website, with attendance numbers added by hand.  Virtually all attendees
in 2012 were therefore recorded as anonymous attendees, for which
neither gender or field of study is known.

While these assignments of gender and field of study used some of the
user's personal attributes in the system, once the assignment was done
the personal information was no longer needed.  The subsequent
analysis of trends was performed having removed names, emails,
institutions, supervisor information, and any other identifying information, using
only anonymous data.

\section{Results}

\begin{figure}[t]
  \includegraphics[width=\columnwidth]{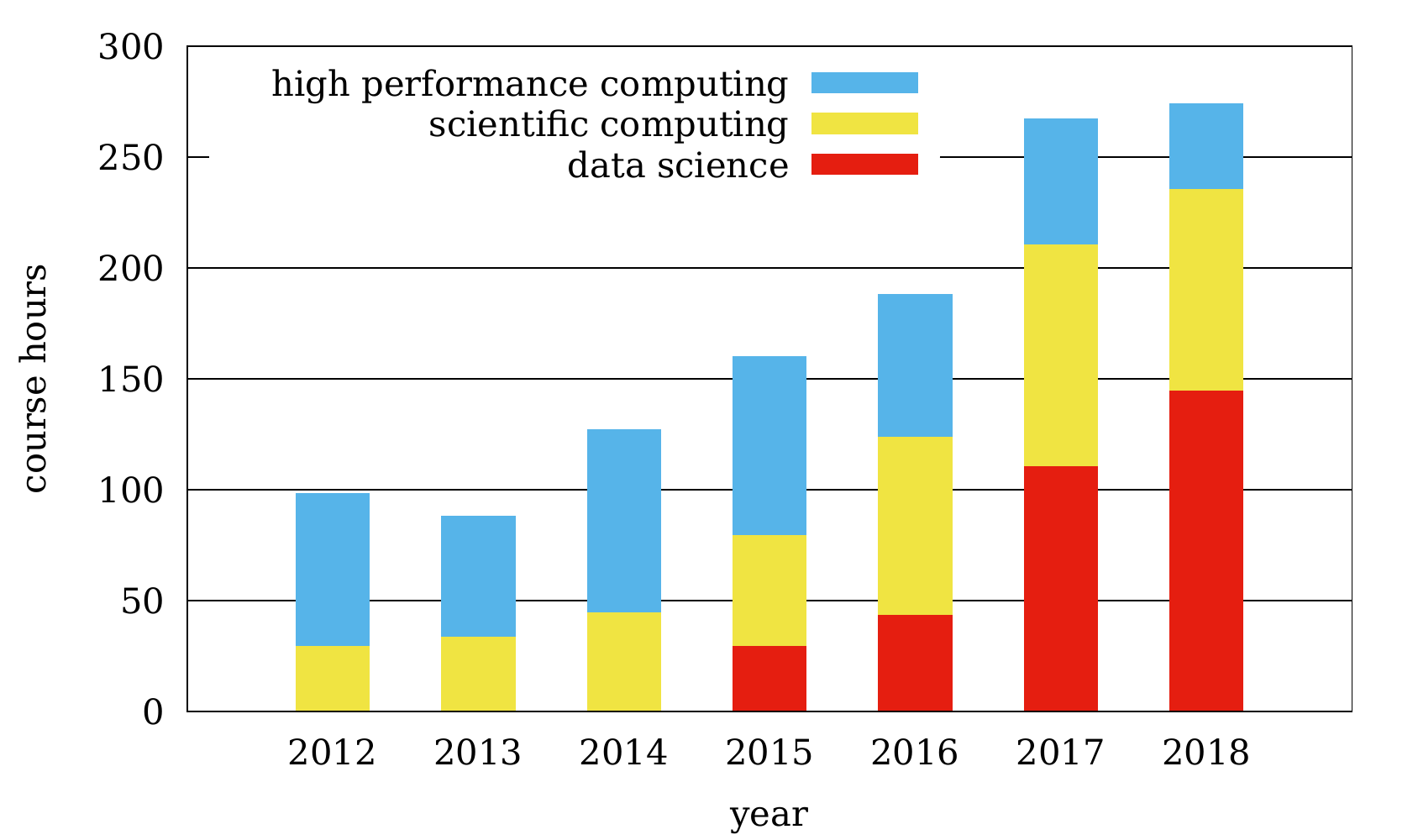}
  \caption{Growth in total course hours delivered by SciNet. The shaded
    regions in each bar show the course hours in the topics scientific
  computing, high performance computing, and data science.\label{fig:coursehoursbytopic}}
\end{figure}

Before presenting the results, two central
notions must be introduced: the first is a course hour,
which is an hour in which an event was held, be it a lecture, seminar,
presentation, meeting, or some other type of event.  The number of
course hours is a measure of the teaching effort.  The second notion
is that of an attendance hour, which is one person attending one hour
of training.  For example, if a training event of two hours has ten
attendees, that training event counts as twenty attendance hours.  Attendance hours
are a measure of the effect of the teaching and to some extent an
indicator of the demand for that training.

\subsection{Overall Growth and Distribution by Topic}

Figure \ref{fig:coursehoursbytopic} shows the overall growth of the
number of hours of training events delivered by SciNet over the
years.  One sees a general increasing trend from about 100 course
hours in 2012 to over 250 course hours in 2018. One can also see a
levelling off of this trend in the last two years.  This can be
attributed mostly to limits in available human resources.

The same figure also shows how many of these training hours were
devoted to the three main categories of topics that are taught: data
science, high performance computing, and scientific computing.  It may
seem that no data science was taught before 2015, but actually the
distinction between data science and scientific computing was not yet
made at that time.  One sees a strong surge of data science training,
which now comprises more that 50\% of SciNet's training effort.

\begin{figure}[t]
  \includegraphics[width=\columnwidth]{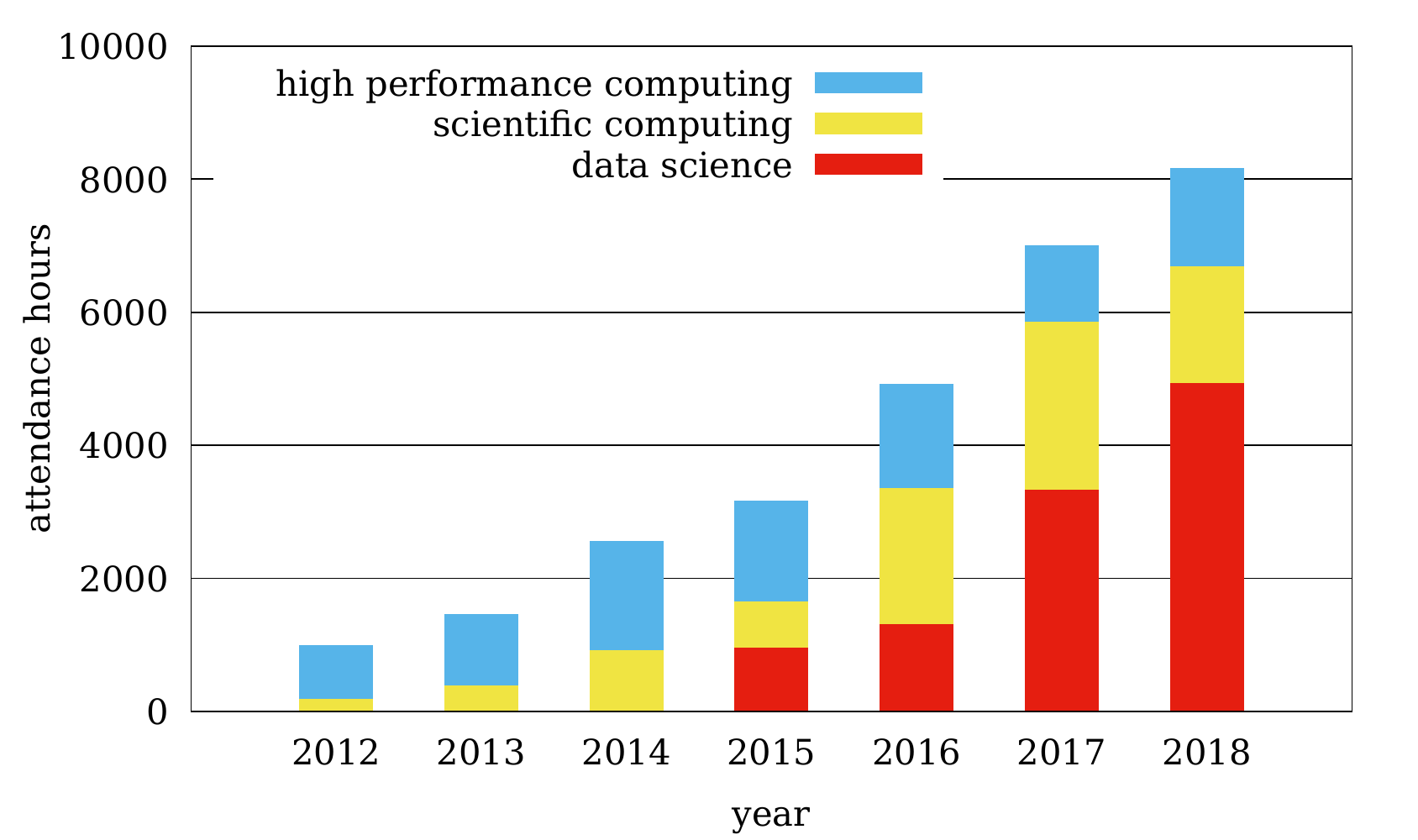}  
  \caption{Break-down of the attendance for the years 2012-2018 by the
    three high-level categories of SciNet courses: scientific
    computing, high performance computing, and data science.
    \label{fig:attendancebytopic}
  } 
\end{figure}

Figure \ref{fig:attendancebytopic} shows the overall growth of the
number of attendance hours over the years.  Once more, one sees a
general increasing trend from about 1000 attendance hours in 2012 to
over 8000 in 2018. The latter number is an underestimate, as the
attendance of our fall classes has not yet been counted. A rough
estimate based on the enrollment numbers suggests that the final
number of attendance hours in 2018 will be closer to 10,000.  There is no sign of attendance levelling off, from which we may
conclude that the demand for training continues to increase.

As in the previous figure, the breakdown by high-level category (data
science, high performance computing, and scientific computing) is also
shown.  Overall, all topics see increasing attendance numbers, but
data science is the fastest growing category,

\subsection{Trends in Participation by Gender and Scientific Field}

Whereas the previous section looked at what is taught, we are also
interested in who is taking SciNet's courses.

\begin{figure}[bt]
  \includegraphics[width=\columnwidth]{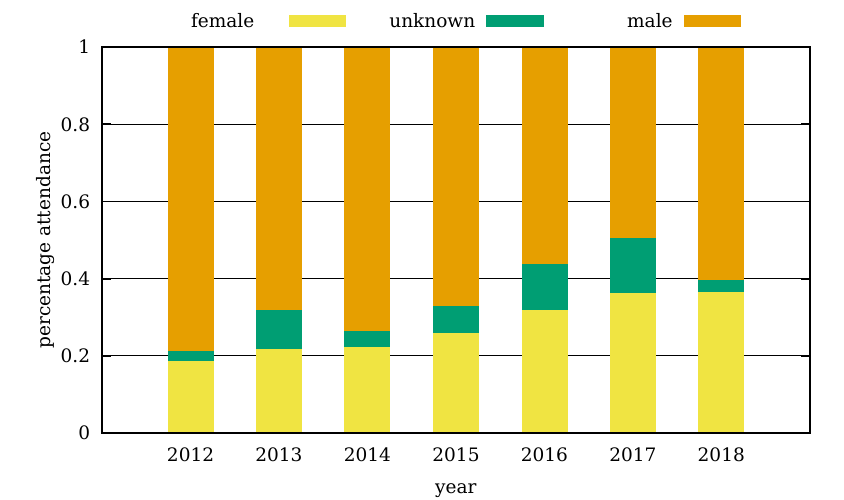}
  \caption{Division of attendance hours in SciNet training by gender (based on an approximate gender determination -- see
    text).\label{attendancebygender}}
\end{figure}

Of particular interest is the gender balance. Figure
\ref{attendancebygender} shows the percentages of different genders.
One must keep in mind that the gender was in many cases inferred rather
than collected. Because of that, it was not possible to go beyond a
basic binary division of genders. The data nonetheless shows a trend
from very little female participation in 2012 to about 40\% female
participation in 2018.

The disciplines that require or use scientific computation are changing
as well, and the training data supports this.  Table
\ref{table:attbyfield} shows the relative participation of students
subdivided into 11 groups corresponding to their field of study.
The largest and second largest groups are highlighted in
orange and yellow in the table.  A very striking trend is apparent
here.  Whereas the majority of students previously came from engineering
and physical sciences, in recent years students in life sciences
(biology and medicine) have become the majority of participants in
SciNet's training.  It is worth mentioning that in absolute numbers,
the engineering and physical science fields have not decreased, but
the life sciences have simply contributed more to the growth in demand
for training.

\begin{table}[b]
  \begin{tabular}{lr@{\quad}r@{\quad}r@{\quad}r@{\quad}r@{\quad}r}
                   & 2013    & 2014    & 2015    & 2016    & 2017    & 2018\\\hline
    engineering    &\cellcolor{orange}\bf34\%  & 14\%    & 15\%    & 20\%    & 12\%    & 17\%\\
    physics        &\cellcolor{yellow}29\%  &\cellcolor{orange}\bf36\%  &\cellcolor{orange}\bf30\%  &\cellcolor{yellow}24\%  & 18\%    & 12\%\\
    earth science  & 10\%    &  8.1\%  &  8.8\%  &  6.0\%  &  3.4\%  &  2.3\%\\
    chemistry      &  6.7\%  &16\%  &  2.7\%  &  4.4\%  &  3.4\%  &  4.4\%\\
    computer science &  1.2\%  &  1.4\%  &  0.3\%  &  1.6\%  &  1.7\%  &  0.8\%\\
    mathematics    &  0\%  &  0\%  &  0\%  &  2.1\%  &  0.8\%  &  2.5\%\\
    medicine       & 10\%    &\cellcolor{yellow} 17\%    & 20\%    &\cellcolor{orange}\bf25\%  &\cellcolor{orange}\bf35\%  &\cellcolor{orange}\bf40\%\\
    biology        &  8.0\%  &  4.8\%  &\cellcolor{yellow}22\%  & 16\%    &\cellcolor{yellow}24\%  &\cellcolor{yellow}18\%\\
    economics      &  0\%  &  0.9\%  &  0\%  &  0.4\%  &  0.4\%  &  2.2\%\\
    social science &  0\%  &  0\%  &  0\%  &  0\%  &  0.2\%  &  0.2\%\\
    humanities     &  0\%  &  0.1\%  &  0\%  &  0\%  &  0\%  &  0\%\\\hline
  \end{tabular}
  \caption{Relative percentage of the attendance in SciNet training by
    researchers from different fields, separated by year. The two
    fields with most participation in a given year are
    highlighted. There was not enough statistics to be able to include
  2012 in this table.\label{table:attbyfield}}
\end{table}

One might expect that the increase in attendance in data science
training is correlated with the increase in attendance from biology
and medicine.  Table \ref{table:catbyfield} shows the division among
scientific computing, data science and high performance computing for
each of the fields of study in the previous table.  This table
confirms what one might already expect: the life sciences are more
concerned with data science, while the physical sciences and
engineering have a greater interest in scientific computing and high
performance computing.

\begin{table}
  \begin{tabular}{lr@{\quad}r@{\quad}r@{\quad}r@{\quad}r@{\quad}r}
                   &  data        &          & scientific \\
                   &  science     & ~~HPC~~  & computing  \\\hline
    engineering    &  20\%   & 46\%    & 34\%    \\
    physics        &  15\%   & 34\%    & 51\%  \\
    earth science  &  18\%   & 41\%    & 41\%\\
    chemistry      &  25\%   & 41\%    & 34\%  \\
    computer science& 12\%   & 72\%    & 16\% \\
    mathematics    &  21\%   & 49\%    & 30\%  \\
    medicine       &  65\%   & 17\%    & 18\%  \\
    biology        &  58\%   & 16\%    & 26\%  \\
    economics      &  26\%   & 56\%    & 18\%  \\
    \hline
  \end{tabular}
  \caption{Relative percentage of participation in training in
  scientific computing (SC), data science (DS) and high performance computing (HPC)
  for different fields of study. There were too few data points for
  sensible results for social science and humanities.\label{table:catbyfield}}
\end{table}

\subsection{Summer School Statistics}

The annual summer school has been highly successful and has been growing in
number of sessions (compare table~\ref{table:summerschool2009} and
table~\ref{table:summerschool2018}) and in attendance.  From 2012 to 2018,
the attendance has grown from 35 people to 215.

After completing at least 3 days of the summer school,
participants in the summer school receive a certificate of attendance. These
are special summer school certificates that are separate from the
SciNet certificates.
In 2012, 20 certificates were awarded, but by 2018, this number was
135.  This growth is partly due to the inclusion of a data
science stream and a biomedical stream.

The school is offered for free, but without support for travel,
lodging or meals. It is therefore not surprising that most
participants are from the Toronto area, although there are always some
who travel to attend the event. In 2018, there was a sizable
number of attendants from outside Toronto (60), from outside of
Ontario (15) and even from outside Canada (5).

This event is in high demand: in 2018, within one day of opening the
registration, there were over 100 registrations, and just one week
later, the 200 registration mark was crossed.  In the end, 304 people
signed up. The fact that there is no charge for this event means that
not everyone attends; attendance rates of 70\% are typical.  This helped the people on the waiting list, most of whom could be accommodated in
the end.

\subsection{SciNet Certificates Statistics}

\begin{figure}[t]
  \includegraphics[width=\columnwidth]{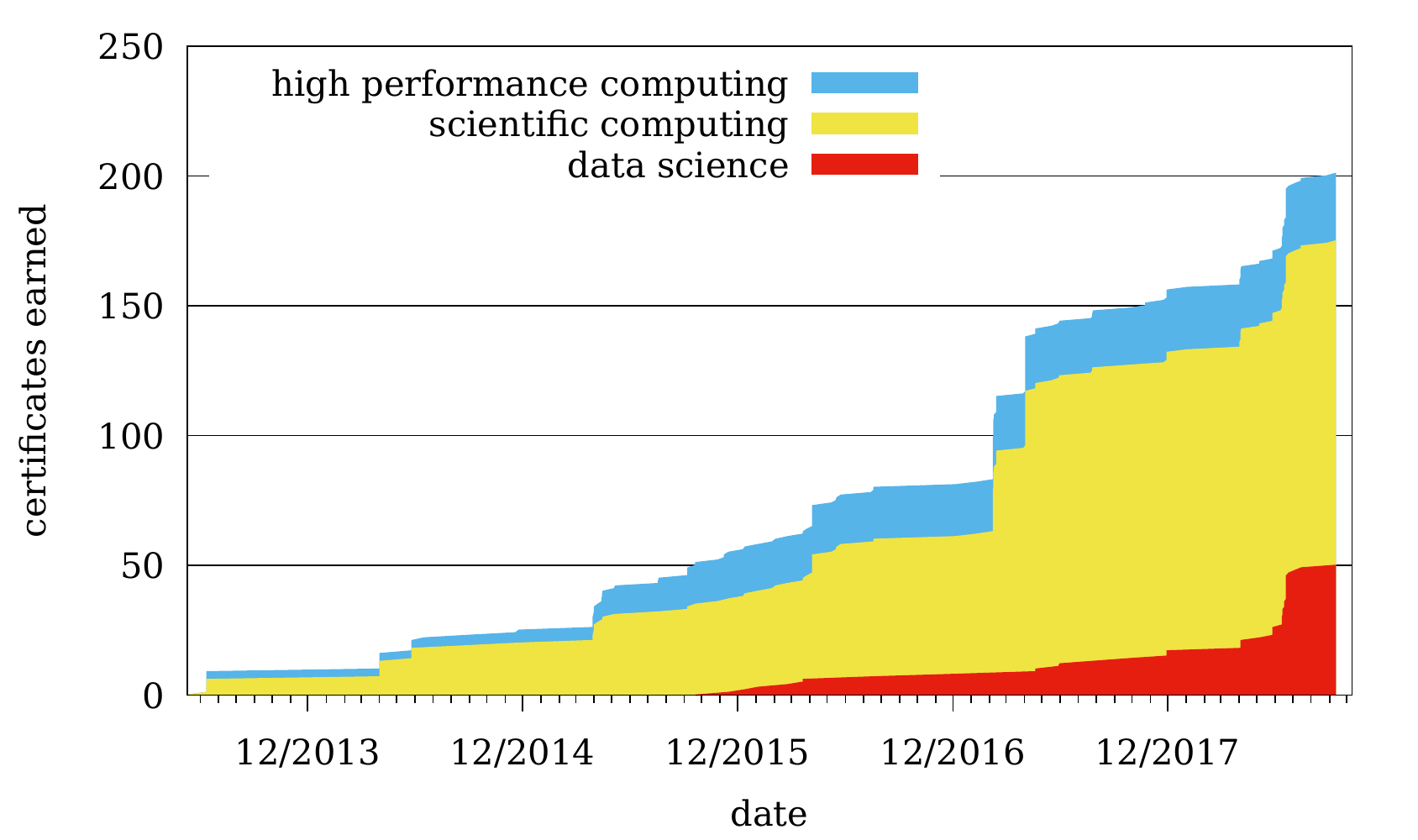}
  \caption{Certificates earned in the last six years (cumulative) in
    each of the categories scientific computing, data science and high performance computing.\label{fig:certificates}}
\end{figure}

As was mentioned above, SciNet issues certificates in scientific
computing, data science, and high performance computing to students
who have taken at least 36 hours of courses in the respective
category.  As figure \ref{fig:certificates} shows, over 200
certificates have been issued so far.  The  data science
certificate still has the lowest number (it was only started in 2015),
but is the fastest growing category.  The graph shows a stagnation
in the number of high performance computing certificates.  It is possible that
this is due to a shift in demand from HPC to data science, but it
could also be related to the decrease in the number of HPC courses
that are on offer; as figure \ref{fig:coursehoursbytopic}
shows, the number of HPC training hours has decreased in the last two
years.

\subsection{Retention}

\begin{figure}[t]
  \includegraphics[width=\columnwidth]{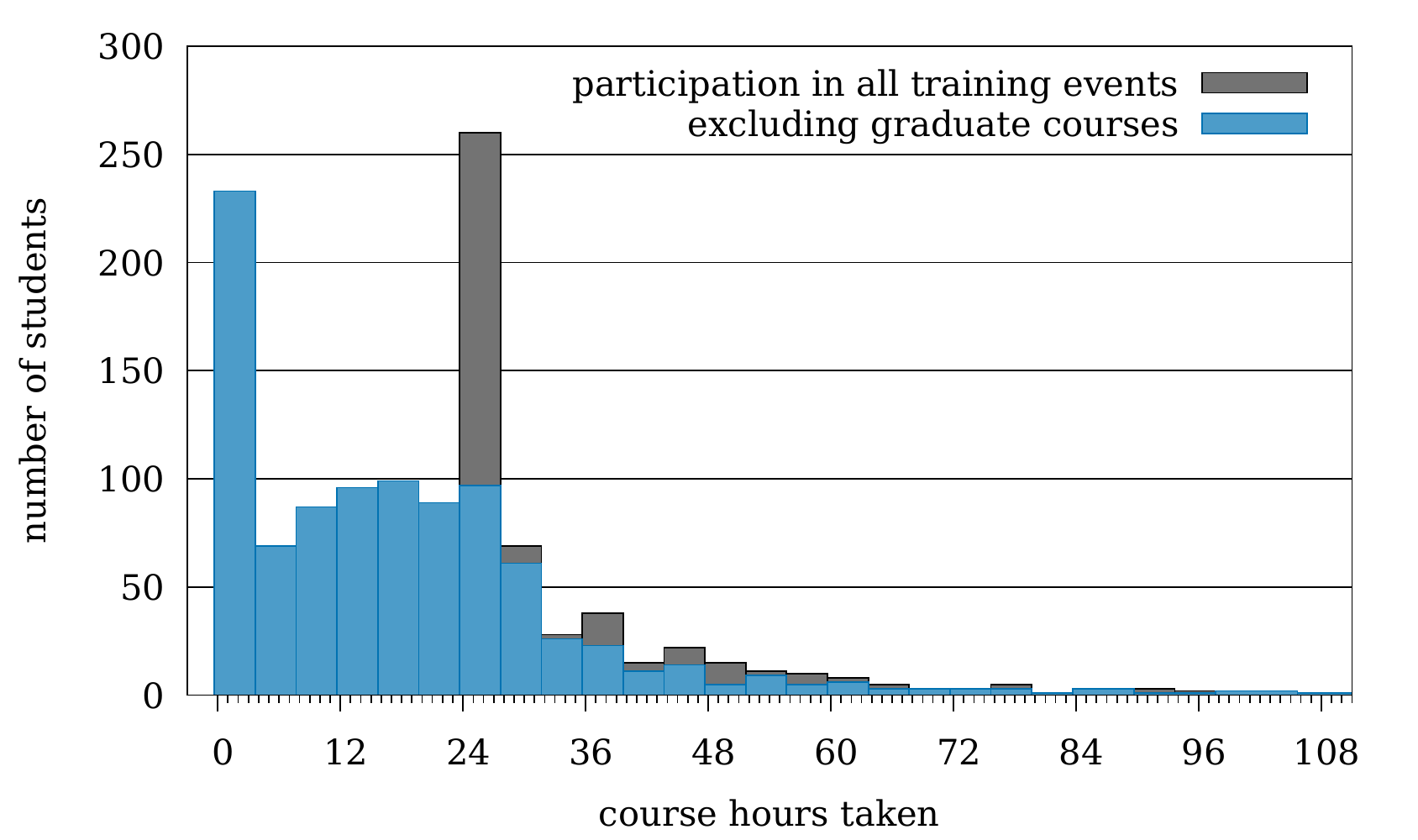}
  \caption{Histogram of the number of course hours taken by
    students. The bin width is 4 hours. Gray bars include all training
    activities, while blue bars exclude for-credit
    courses.\label{fig:hourhist}}
\end{figure}

The statistics of the number of certificates give an indication of the
retention of students taking SciNet courses, as one course is not
enough to earn a certificate.  We can get more insight into the
retention rates by considering the number of course-hours taken by
individuals.  Figure \ref{fig:hourhist} shows the number of students
that spent a given number of hours in SciNet training, counted in bins
of 4 hours wide (i.e., the first bin counts students that spent
between 1 and 3 hours in SciNet courses, the second counts students
that took between 4 and 7 hours, etc.).  It also shows those same
statistics when the graduate courses are taken out.

A few things can be observed.  One is that a substantial number of
attendees (about 230/1800) come to only one to three hours of
training.  Since there are very few individual training events with a
duration of more than three hours, the remaining attendees are
returning students.  The second observation is that there is a large
peak at 24 hours of attendence.  This peak disappears when the
graduate courses are taken out, which makes sense as all our graduate
courses have 24 hours of lectures. What then remains is a broad
distribution, centered around 16 hours.  There is also a noticeably
long tail, which shows that a number of students attend a lot of
training.

\section{Discussion}

In this paper, we have shown the growth and demand-driven change in
focus of SciNet's training program, from the year 2012 to 2018.  In
the analysis of the enrollment data, we had to infer gender and field
of study (we are exploring ways of implementing ways for users to
self-identified their gender and field of study). The data shows a
large increase in life science participation as well as in female
participation. There has been an overall growth trend both in the
amount of training offered and in the attendance. In particular,
training in the field of data science has shown remarkable growth.
The tremendous increase in participation shows the large demand for
this kind of training.

One might wonder whether or not this kind of training should be part
of the graduate curriculum \cite{SC.HPC.HigherEd-inpress}, or whether
perhaps it should be provided by the Computer Science department.
While parallel computing and concurrency are active lines of research
in computer science, users are often interested in practical
techniques and desire training in efficient ways to enable larger
scientific computations.  Other departments might want to teach
courses on data science and scientific computation, but may not have
the required knowledge.  Computing centers usually have the required
expertise, and can provide much of the necessary training, but may not
be in a position to give or create for-credit university courses.  At
least in the case of SciNet, through partnering with other
departments, several graduate courses in Scientific Computing have
been developed.

Overall, this has been a successful program which continues to
grow.

\section*{Acknowledgements}
SciNet is funded by the Canada Foundation for Innovation, the
Government of Ontario, and the University of Toronto.


\bibliographystyle{ACM-Reference-Format}
\bibliography{refs}

\end{document}